\documentclass[10pt]{article}  
\usepackage{cite}
\usepackage{amsmath,amssymb}
\usepackage{tabularx}
\usepackage{subcaption}
\usepackage{graphicx}
\usepackage{url}

\begin{document}

\title{\bf Measurements of Radiation Pressure owing to the Grating Momentum}

\author{Ying-Ju Lucy Chu$^1$, Eric M. Jansson$^2$, and Grover A. Swartzlander, Jr.$^{1*}$}

\date{\small
$^1$Chester F. Carlson Center for Imaging Science, Rochester Institute of Technology\\
$^2$Charter School of Wilmington, Wilmington, Delaware\\
\today}

\maketitle

\begin{abstract}
The force from radiation pressure owing to the grating momentum was measured for
a thin transmissive fused silica grating near the Littrow angles at wavelengths of 808 nm and 447 nm.
A significant magnitude of force was measured in the direction parallel to the grating surface.
We also confirmed that the component of force normal to the grating surface may vanish.
This forcing law is characteristically different from radiation pressure on a reflective 
surface, and thus, opens new opportunities for light-driven applications such as solar 
or laser driven sailcraft, or the transport of objects in liquids.
\end{abstract}
Since Maxwell first predicted radiation pressure in 1873 \cite{maxwell},
it has helped to describe phenomena ranging from the astronomical to the quantum realm.
For example the gravitational collapse of stars and accretion dynamics are
governed by radiation pressure
\cite{schwarzschild1901druck,eddington1918stars}.
Experimental evidence of Kepler's 1619 explanation of comet tails
\cite{lebedev1901investigations,nichols1903application}
was later extended to the general distribution of interplanetary dust  \cite{gindilis1969solar,schwehm1976radiation}.
Terrestial applications have found uses in biology as optical tweezers \cite{ashkin1980applications}, laser cooling of atoms\cite{lett1988observation,dalibard1989laser}
and macroscopic objects \cite{metzger2004cavity,bhattacharya2017levitated}. 
The detection of gravitational waves by means of laser interferometers requires an accounting of radiation pressure \cite{corbitt2006measurement}. 
Micro-structures such as optical wings \cite{swartzlander2011stable}
and slot waveguides have promising photonic applications \cite{yang2009SlotWaveguide,liu2013highly}.
Thin microfabricated sheets such a diffraction gratings and diffractive metamaterials \cite{stork1991artificial,oh2008achromatic,lavrentovich2011liquid,marrucci2013q,zhang2015ultra,tabiryan2015thin,gupta2016single}
provide opportunities to marry recent developments in materials research 
with grand ambitions for astronautical space travel.
For example, radiation pressure is one of the few methods of reaching distant stars with free sunlight \cite{frisbee2009limits,gilster2004centauri}
or expensive laser systems \cite{marx1966interstellar,forward1984roundtrip}. 
While those sailcraft considered elementary attitude-controlled reflective sails, optical scientists have recently proposed 
passive or active diffractive sails that may provide superior control authority for near-Earth missions and beyond
\cite{groverdiffractive, achouri2017metasurface}, 
owing to force components along both the surface normal and tangent.  Unlike a reflective sail
that has only a normal component of force, a diffractive sail has both a tangential and a normal
component of force. The latter is notable for changing sign, continuously passing through the zero-value point
as the angle of incidence is varied.
The large tangential component of force of a diffractive sail may be 
particularly advantageous for raising or lowering the orbit of a sailcraft\cite{groverdiffractive}.
The experimental measurements described below validate the premise that diffraction gratings experience 
wavelength-dependent force components in both the normal and transverse directions.

Although the magnitude of radiation pressure may seem relatively weak
owing to its inverse relation to the speed of light,
this value may be comparable to the gravitational force in outer space or in a
quasi-neutrally buoyant liquid. 
This provides astronautical opportunities to propel low-mass sailcraft through space
and a new laboratory technique to assert non-contact forces on small bodies. 
Light-driven sails being developed for future space travel afford cheap
and inexhaustible energy for a myriad of missions 
\cite{mcinnes2013SolarSail,garner1999summary, johnson2011nanosail}.  
Similar to the development of air flight in the early 1900's, sailcraft technology is likely 
to rapidly advance after in-space demonstrations reveal the extent of navigation-by-light challenges.
New materials and sailcraft architectures will be perfected to optimize particular mission 
objectives. For example, one may question the necessity 
of a reflective film that transfers electromagnetic momentum to mechanical momentum.  
A diffractive film provides advantages over a reflective film.
For example, a micrometer-scale reflective film must be tilted to change the direction 
of force -- a daunting task if the sail area is hundreds of square meters.  
However, non-mechanical beam steering of a 
diffractive element (e.g., using electro-optic techniques) 
overcomes this complication \cite{tsuda2013achievement}.  Although the concept of a grating 
momentum vector is commonly invoked to predict the direction of transmitted and reflected 
beams from a grating, to our knowledge, the radiation pressure force on diffraction 
grating plate has not been measured.  Here we report our experimental investigation of
the relationship between the grating momentum and radiation pressure force on a 
commercial grating.  To our knowledge this is the first time such values have been measured.

The radiation pressure force on a grating may be expressed
\begin{equation}
\vec{F}=\frac{P_i}{ck}\left(\vec{k}_i-\sum\limits_{j}\eta_j\vec{k}_j \right)
\label{eq:radiation_force}
\end{equation}
\noindent where $\eta_j = P_j / P_i$ is the efficiency of the $j^{th}$ diffracted beam,
$P_i$ ($P_j$) is the incident (diffracted) beam power, 
$c$ is the speed of light, 
$\vec{k}_i$ ($\vec{k}_j$)
is the incident (diffracted) wave vector, with $k = |\vec{k}_i| = |\vec{k}_j| = 2 \pi / \lambda$,
and $\lambda$ is the wavelength of the beam of light.  
Doppler shifts caused by the relative velocity of the grating are ignored here.

\begin{figure}[h!]
\begin{center}
\includegraphics[width=0.50\linewidth]{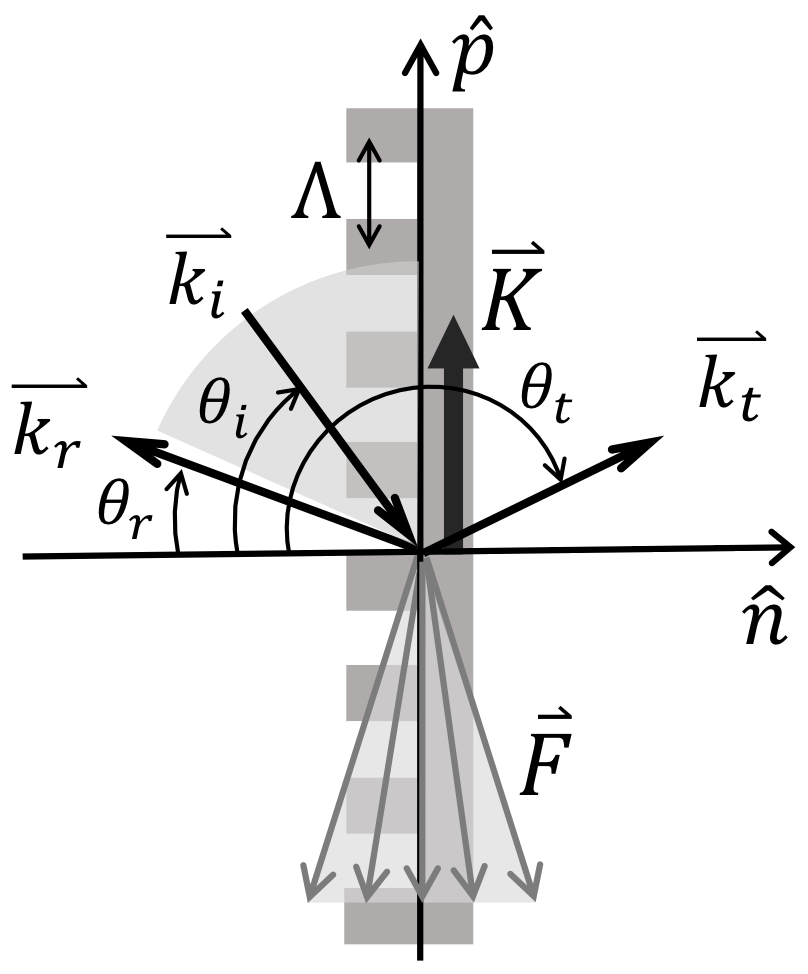}
\end{center}
\captionsetup{margin=10pt,justification=raggedright,singlelinecheck=false}
\caption[System coordinates.]
{Plane of incidence for a  diffraction grating of period $\Lambda$, with respective incident,
reflected, and transmitted angles $\theta_i$ , $\theta_r$ , $\theta_t$, 
wave vectors $\vec{k}_i$ , $\vec{k}_r$ , $\vec{k}_t$, and grating momentum $\vec{K}=(2\pi/\Lambda) \hat{p}$. 
The allowed range of incident angles, $\sin^{-1}(m\lambda/\Lambda-1)$ to $90^{\circ}$, 
produces a force $\vec{F}$ having a constant tangential component, $F_p$ and a
positive or negative normal component, $F_n$.}
\label{grating_angle_define}
\end{figure}

In this report we assume negligible absorption, convective, photophoretic, and outgassing forces.
Further, the forcing laser beam is assumed to under-fill the grating surface.
A depiction of incident and diffracted beams for a single diffraction order grating, 
with corresponding angles,  $\theta_i$, $\theta_t$, and $\theta_r$, is shown in 
Fig. 1.  
Phase-matching of the electromagnetic fields at the grating boundary provides a 
relation between the components of the wave vectors that are parallel to the surface:
\begin{equation}
\vec{k}_i \cdot \hat{p} + m \vec{K} = \vec{k}_m \cdot \hat{p}
\label{eq:k_vectors}
\end{equation}
\noindent where $\vec{K} = (2 \pi / \Lambda ) \hat{p}$ 
is the so-called grating momentum (the factor $\hbar$ is
typically ignored), $\Lambda$ is the grating period,
$\vec{k}_m$ is the $m^{th}$ diffraction order (for either the reflected
or transmitted beam), and $\hat{p}$ ($\hat{n}$) is the 
unit vector parallel (normal) to the grating surface.  
The well-known grating equation is a restatement of 
Eq. (2): 
$ \sin\theta_m = -\sin\theta_i + m \lambda / \Lambda $.


For an ideal grating with unity transfer efficiency into a single diffraction order
the parallel and normal force components may be expressed 
\begin{subequations}
\begin{align}
F_p& = -(P/c)(m\lambda / \Lambda ) \\
F_n& = (P/c)(cos\theta_i \pm (1 - (m\lambda / \Lambda - sin\theta_i)^2)^{1/2})
\label{eq:Fp_Fn_efficiency}
\end{align}
\end{subequations}

\noindent 
where the minus (plus) sign is for a transmissive (reflective) diffraction order.
The force efficiency components, $F_p c/P_i$ and $F_n c/P_i$ of a transmissive grating
are plotted in Fig. 2
for the allowed diffraction angle(s) at two wavelengths corresponding to our 
experiments: $\lambda =$ 808 nm and 447 nm.
The parallel component of force is negative, as expected from 
conservation of momentum arguments (see Fig. 1).  What is more, $F_p$
is independent of the incident angle $\theta_i $ when 
$| m \lambda / \Lambda -1 | < 1$ is satisfied.
In contrast, the normal component of force is positive below the 
Littrow diffraction angle, defined by $2 sin\theta_{i,L} = m \lambda/ \Lambda$.
For $\theta_i > \theta_{i,L}$ the normal component of force is negative
and the light source acts as a partial "tractor beam" \cite{novitsky2011single,sukhov2011negative,sukhov2010concept,palima2013optical},
pulling rather than pushing the grating.
The normal force components vanish at the Littrow diffraction angle (marked as diamonds in Fig. 2). 
Light sailing and terrestrial applications inspired us to measure the components of force on a diffraction grating
with a torsion oscillator \cite{gillies1993torsion}.

\begin{figure}[h!]
\centering
\includegraphics[width=1\linewidth]{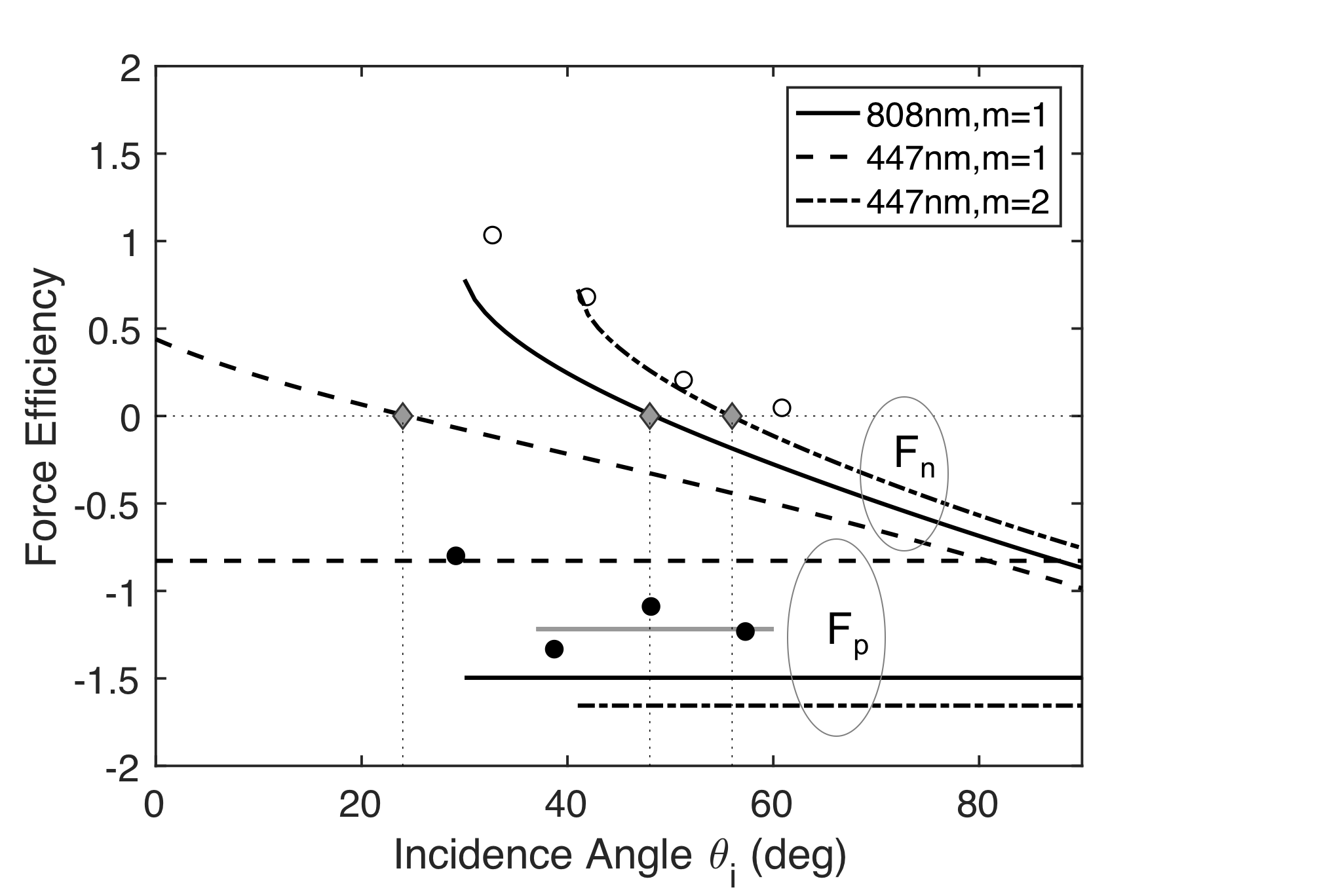}
\captionsetup{margin=10pt,justification=raggedright,singlelinecheck=false}
\caption[Ideal force efficiency and experimental values.]
{Parallel and normal force efficiencies, $F_p c / P_i$ and $F_n c / P_i$ respectively,
for an ideal single order transmission diffraction grating
with order $m$, wavelengths $\lambda=$ 808 nm and 447 nm, and grating period $\Lambda = 540$ nm. 
Normal force efficiencies vanish at the Littrow angles (diamond points): 
$48.4^{\circ}$ (808 nm, m=1), $24^{\circ}$ (447 nm, m=1), $56^{\circ}$ (447 nm, m=2). 
Experimentally determined values (circles) are shown for $\lambda=808$ nm.}
\label{fig:force_efficiency}
\end{figure}

The force on a non-ideal non-absorbing grating must account for
multiple diffraction orders.
In this case the parallel and normal components of force 
may be respectively expressed
\begin{subequations}
\begin{align}
F_p& = - \frac{P_i}{c}\sum \left[
\eta_{j,r} (\sin\theta_i + \sin\theta_{j,r})  + 
\eta_{j,t} (\sin\theta_i + \sin\theta_{j,t})\right] \\
F_n& = \frac{P_i}{c}\sum \left[
\eta_{j,r}(\cos\theta_i + \cos\theta_{j,r}) + 
\eta_{j,t}(\cos\theta_i + \cos\theta_{j,t})\right] 
\label{eq:Fp_Fn}
\end{align}
\end{subequations}
\noindent 
In principle, Eq. (4b) may also allow a zero-valued normal force component,
resulting in a purely tangential force, as described above.

We built a torsion oscillator using a $D = $25 $\mu$m diameter, $L_f$ =240 mm long tungsten filament 
(Alfa Aesar 10405-H4). It was attached to an aluminum support frame at one end,
and a suspended twist-hardened copper wire of length $2R =$ 220 mm, diameter 1 mm, 
at the other end.
An optimized single order diffraction grating (LightSmyth T-1850-800s-3210-93) 
having a period $\Lambda$ = 540 nm 
was ground to a thickness of 190 $\mu$m and attached in one of two
configurations:  (A) with its surface normal parallel to the copper wire;
(B) with its surface normal perpendicular to the copper wire (see Fig. 3).
A balancing mass was placed on the other end of the wire.   
A small lightweight mirror was attached at the vertex
of the wire and filament to allow measurements of the angular displacement, $2 \delta \approx S/L$, of a low power
HeNe tracking laser,
where $S$ is the linear displacement of the laser beam from its equilibrium position on a screen
placed a distance $L = 1.92$ m from the pivot.
Time lapse photographs (Canon 5D III and Canon TC-80N3) of the screen were recorded 
at $\Delta t = 4 $ s intervals.  
The position of the beam was obtained by determining the beam centroid in each image.
The apparatus was transported to a suburban basement that was remarkably free of characteristic vibrations and 
loading sag experienced on our institutional building bedrock floor.  An aluminum wire mesh was shaped into a $300$ mm 
high cylinder to serve as a Faraday cage, shielding the oscillator from inadvertent electrostatic torques. 
The system was centered within a customized borosilicate bell jar of good optical quality.
After evacuating air from within the bell jar to a pressure of $~10^{-5}$ hPa ($7.5\times 10^{-6}$ Torr), 
the disturbed oscillator was brought to near rest by means of radiation pressure from the forcing laser.
At this pressure the mean free path of the remaining air molecules exceeded the diameter of the bell jar.
The system remained at rest for many hours -- even while the vacuum system labored and people walked nearby.
We attributed this stability to concrete-on-earth flooring.

The measured period of free oscillation was $T_0$ = 100.8 s,
and the characteristic decay time ($1/\alpha$) was roughly $80 T_0$.
Based on the calculated moment of inertia, $I = 1.0\times 10^{-5} $ kg$\cdot$m$^2$,
and the measured period, the torsional spring constant of the filament was determined to be  
$\kappa = (2 \pi / T_0)^2 I = 3.9\times 10^{-8}$  N$\cdot$m/rad.
This value agreed well with the theoretical value obtained from a tabulated value 
of the Young's modulus $Y = 410$ kN$\cdot$mm$^{-2}$ \cite{the_engineering_toolbox}:
$\kappa ' = \pi Y D^4 / 32 L_f=6.6 \times 10^{-8}$ N$\cdot$m/rad.  The discrepancy
between the two values is attributed to the unknown value of $Y$ for thin filaments 
of unknown working history,
and to the unknown variability of $D$ along the length of the filament.
We note that the radiation pressure force was not expected to induce significant
linear pendular displacements of the hanging $M=2.4$ g mass.
Several types of measurements are reported below.
 
\begin{figure}[h!]
\begin{center}		\includegraphics[width=1\linewidth]{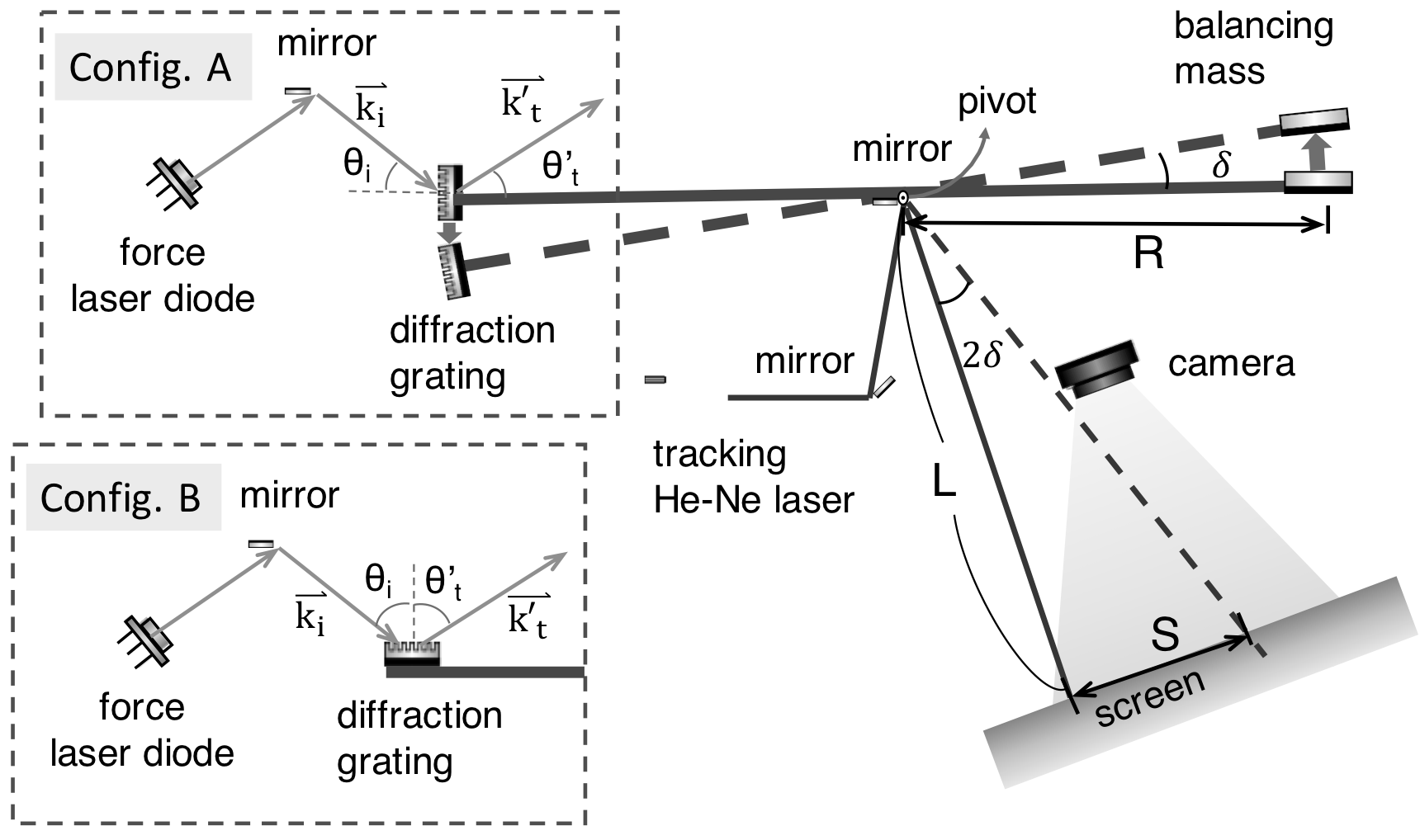}     
\label{fig:pendulum}
\captionsetup{margin=10pt,justification=raggedright,singlelinecheck=false}
    \caption [Schematic of the torsion oscillator.]
    {Top view schematic. Torsion oscillator with moment arm of length $R$, 
    angular displacement $\delta$, forcing laser, tracking laser, camera, screen,
    and diffraction grating in Configuration A or B.}
\end{center}
\end{figure}

\begin{figure}[h!]
    \begin{subfigure}{0.48\linewidth}
		\includegraphics[width=1\textwidth]{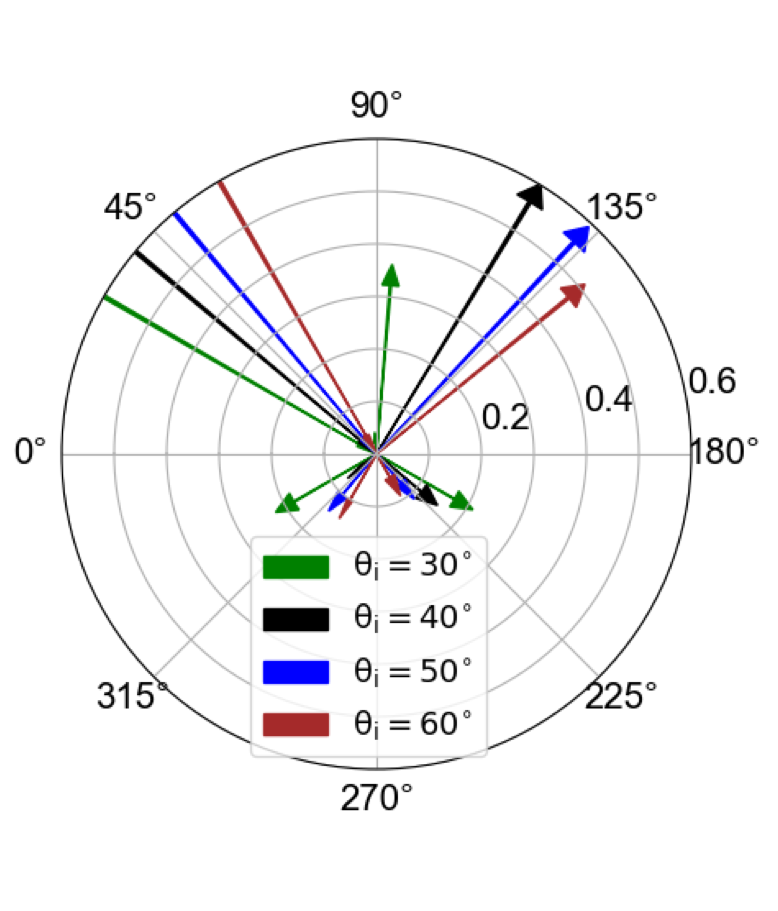}   
        \label{fig:808nm_a}
        \caption{}
    \end{subfigure}
    ~
    \begin{subfigure}{0.48\linewidth}
		\includegraphics[width=1\textwidth]{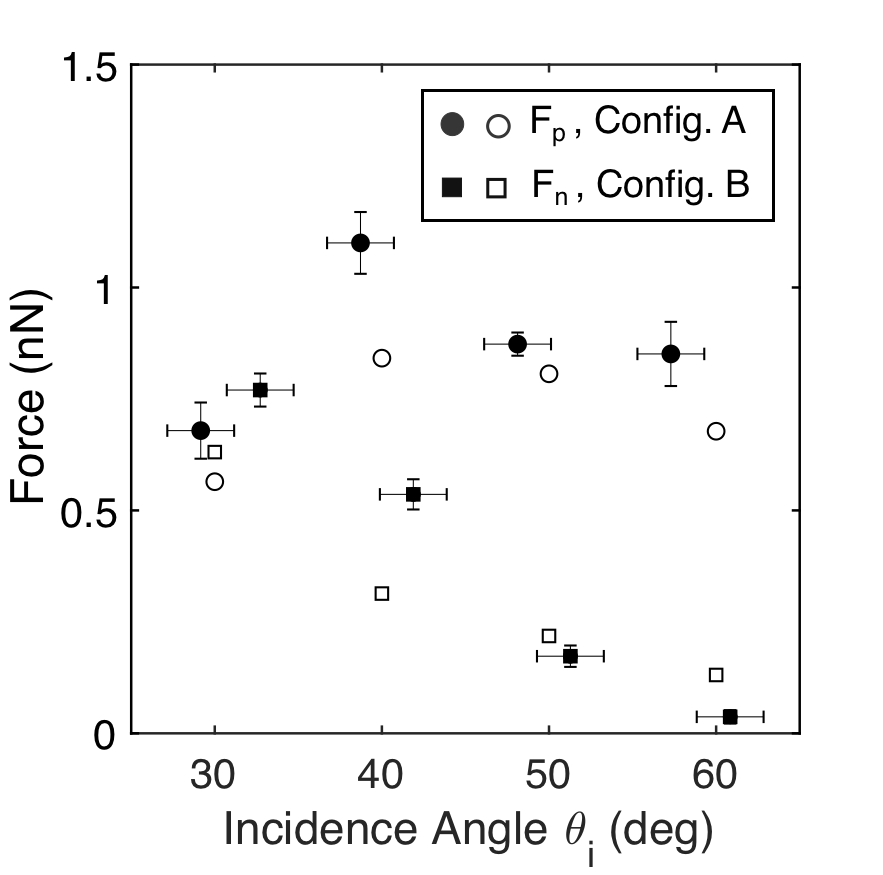} 
        \caption{}
    \end{subfigure}
\captionsetup{margin=10pt,justification=raggedright,singlelinecheck=false}
    \caption[Measurements at 808 nm.]
    {Radiation pressure at $\lambda=$ 808 nm, $P_0 = 345$ mW.
    (a) Measured efficiencies of reflected and transmitted beam powers for
    four incident angles. Grating surface (not shown) aligned along
   $90^\circ$ - $270^\circ$ line.
   (b) Measured (solid points) and predicted (hollow points)
   magnitudes of $F_p$ and $F_n$.}
\end{figure}

First we mounted the diffraction grating with its surface normal parallel to the 
copper torsion arm, as depicted in Fig. 3, configuration A.
The grating lines were transverse to the plane of incidence.
With the bell jar removed, the oscillator was immobilized to allow measurements of the 
transmitted, diffracted, and reflected beams with the forcing laser 
($\lambda =$ 808 nm, and linear polarization transverse to the plane of incidence).
The measured diffraction efficiencies are depicted in Fig. 4(a) for four different angles of incidence: 
$\theta_i=30^{\circ}$, $40^{\circ}$, $50^{\circ}$, $60^{\circ}$.  
The transmitted first order diffraction 
efficiency was expected to be optimal near the Littrow angle $\theta_i=48.4^{\circ}$.  
In fact both the $40^{\circ}$ and 
$50^{\circ}$ beam angles provided peak diffraction efficiencies of roughly $60 \%$.  
We believe $\sim 18 \%$ of the beam power was lost owing to scattering attributed to
micro scratches from the mechanical thinning process (see Table 1).
Although the force from directional scattering may be small but non-zero,
we were unable to measure the angular scattering spectrum; hence it is ignored in our analysis.
To account for Fresnel reflections at the outer and inner diameters of the borosilicate bell jar
we reduce the measured incident power with the angle-dependent, wavelength-dependent transmission factors $T$ listed in Table 1.
The effective power incident on the diffraction grating may be expressed 
$P_i = (1 - P_s / P_0) T P_0$, where $P_s$ is the scattered power. 
The theoretically expected values of force (Eq. (4)) based on the effective incidence power are plotted in Fig. 4(b) (hollow circular data points) for a laser output power of $P_0=345$ mW.   
For example, we calculate $| F_p | \sim 10^{-9}$ N, with angular variations attributed 
to the diffraction efficiency of the grating and Fresnel coefficients of the bell jar.

\begin{table}[h!]
\centering
\setlength{\tabcolsep}{3pt}
\renewcommand{\arraystretch}{1.3}
\begin{tabular}{|l|llllll|}
\hline
$\lambda$ = 808 nm,  $\theta_i$ & 30$^{\circ}$ & 40$^{\circ}$ & 50$^{\circ}$ & 60$^{\circ}$ &  &  \\ \hline
Config. A, $T$ & 0.89 & 0.87 & 0.83 & 0.78 &  &  \\
Config. B, $T$ & 0.78 & 0.83 & 0.87 & 0.89 &  &  \\ \hline 
$P_s$/$P_0$    & 0.17 & 0.19 & 0.13 & 0.23 &  &  \\ \hline
\hline
$\lambda$ = 447 nm, $\theta_i$ & 15$^{\circ}$ & 25$^{\circ}$ & 35$^{\circ}$ & 45$^{\circ}$ & 55$^{\circ}$ & 65$^{\circ}$ \\ \hline
Config. A, $T$ & 0.9 & 0.9 & 0.88 & 0.85 & 0.8 & 0.74 \\ \hline
$P_s$/$P_0$    & 0.21 & 0.33& 0.36& 0.23& 0.29& 0.27 \\ \hline
\end{tabular}
\caption{Calculated Fresnel transmission values (bell jar), $T$, and 
measured scattering fraction (grating) $P_s/P_0$.}
\label{Fresnel_table}
\end{table}

Next we enclosed the oscillator within the bell jar, evacuated the chamber, and brought the free oscillator to near standstill. The forcing laser power was set to $P_0=345$ mW and a mechanical shutter was
opened at time $t_0$ to provide a step function force on the grating.  This procedure was repeated three times for 
each of the four incidence angles described above.  The time-varying angular displacement of the tracking
laser upon the screen was extracted and fitted to a well-known equation for a weakly damped harmonic 
oscillator (see Appendix).  Fitting was achieved by varying $t_0$ and the strength of the step function force.
Small amplitude pre-exposure oscillations were similarly fitted to obtain the state of the oscillator at $t_0$.
The excellent agreement between the experimental data and the theoretical model 
with the fitted force value
(typical RMS error $\sim 0.08\%$) confirms the veracity of the model.

The magnitude of the tangential force $| F_p |$ determined by use of 
configuration A are plotted in Fig. 4(b) (solid circles with error bars).
We find good agreement between the determined and predicted values of force.
The force efficiency values, $F_p c / P_i$ are represented by solid circular 
data points in Fig. 2.  For an ideal single order grating,
these values are expected to be constant.  Although our grating was
not ideal, the measured efficiency values were nearly constant over
a range of angles (see solid gray line in Fig. 2), differing in magnitude by a factor of $80\%$.  
This difference between the ideal and measured values is attributed to the additional
reflected and transmitted diffraction orders in our experiment (see Fig. 4(a)).

To obtain experimental values of the normal component of force we changed the orientation of the diffraction grating 
to Configuration B (see Fig. 3).  As described above, the values of force were determined from measured
values of angular displacement.  As shown in Fig. 4(b) these values are comparable to those predicted from 
Eq. (4).  The normal component of force vanishes at $\theta_i \sim 60^\circ$, which is
$\sim 12^\circ$ greater than the Littrow angle.  Again this difference is attributed
to the use of a non-ideal grating.  The force efficiency values, $F_n c / P_i $ are 
represented by open circular data points in Fig. 2, showing larger values than expected
owing to the reflected diffraction order (see Fig. 4(a)).  Unlike the constant value
of efficiency for the $F_p$ component of force, the efficiency for the normal component
decreases with increasing angle.  One may expect the value to become negative for our
experiment for angles greater than $60^\circ$.

\begin{figure}[h!]
\begin{center}
    \begin{subfigure}{0.48\linewidth}
		\includegraphics[width=1\textwidth]{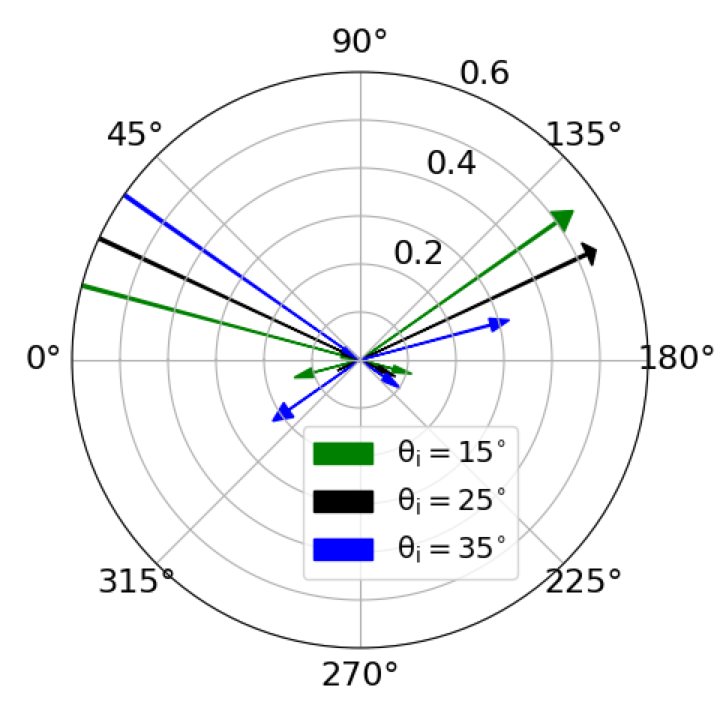}   
        \caption{}
    \end{subfigure}
    ~
    \begin{subfigure}{0.48\linewidth}
		\includegraphics[width=1\textwidth]{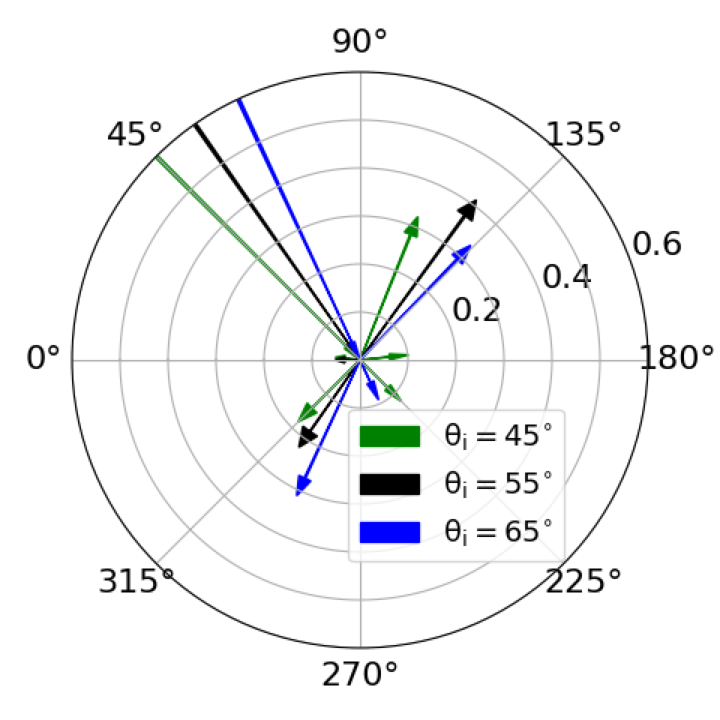}       
        \caption{}
    \end{subfigure}
        \begin{subfigure}{0.75\linewidth}
		\includegraphics[width=1\textwidth]{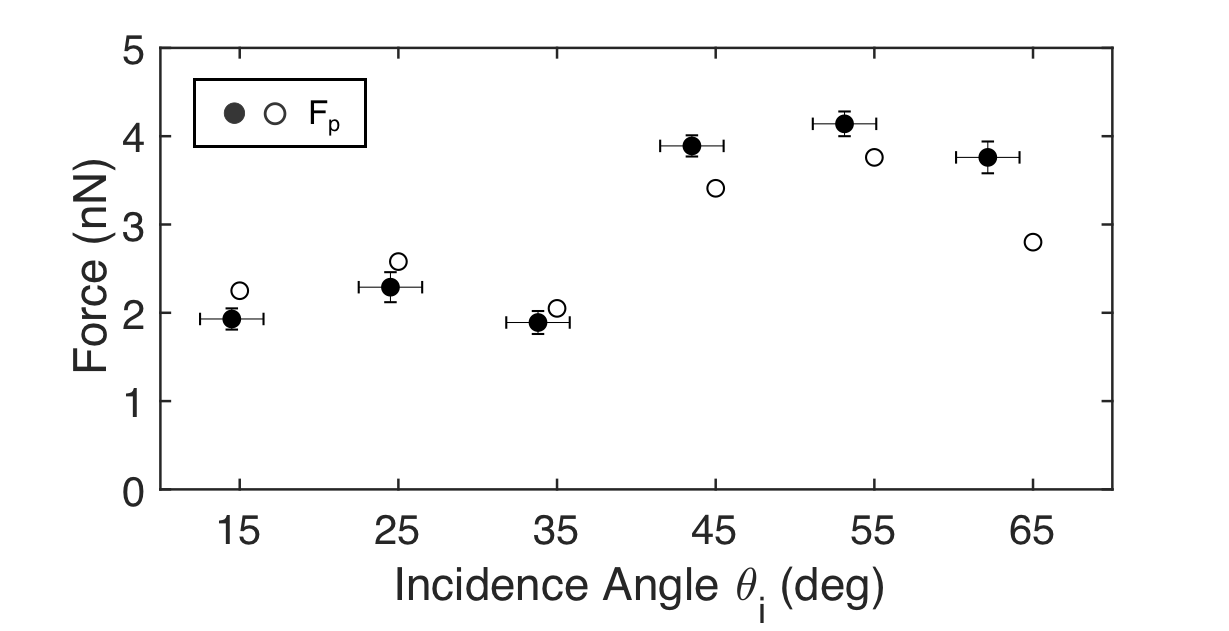}       
        \caption{}
    \end{subfigure}
\end{center}
\captionsetup{margin=20pt,justification=raggedright,singlelinecheck=false}
    \caption[Measurements at 447 nm.]
    {Radiation pressure at $\lambda = 447$ nm, $P_0 = 1.5$ mW. 
    (a,b) Diffraction efficiencies near the first
	and second order Littrow angles, $24^{\circ}$ and $56^{\circ}$ respectively.
    (c) Measured (solid points) and predicted (hollow points) magnitudes of $F_p$.}
\end{figure}

As the wavelength of light changes, the grating efficiency and number of orders may change.
To assess how radiation pressure changes when the grating allows two
diffraction orders, we substituted a laser having a wavelength $\lambda = 447$ nm.  
The procedures described above for Configuration A were repeated.
Measurements were not made in Configuration B.
At this wavelength two Littrow angles are allowed:  $\theta_i = 24^{\circ}$ for $m=1$ and $56^{\circ}$ for $m=2$. 
The measured diffraction efficiency of each transmitted and reflected diffracted order is depicted in Fig.5(a,b),
and the values of $| F_p |$ are plotted in Fig. 5(c) for a laser output power $P_0 = 1.5$ W.
As expected the magnitudes of force are relatively constant, with the 
second order values roughly twice the first order values, as expected (see Fig. 2). 
The average measured force efficiencies, 0.67 for $m=1$ and 1.34 for $m=2$
differ by factor of 2.0, as expected.  
Although the grating was not designed for use at $\lambda = 447$ nm, the
measured values of efficiency agree remarkably well ($84\%$) with the ideal case.

In summary, we have used a vacuum torsion oscillator in two configurations and
at two different wavelengths to measure the radiation pressure force on a diffraction grating.
We have verified that unlike an ideal flat reflective surface,
which is known to have only a normal component of force,
a diffraction grating generally experiences
radiation pressure force components that are perpendicular to and parallel to 
the surface.  The tangential component is
a direct mechanical manifestation of the grating momentum vector. 
As expected, we found that the latter force is roughly constant
and proportional to the grating momentum, whereas the normal
component varied with the angle of incidence.  The normal component
of force vanished, producing a purely tangential force when $\theta_i \sim 60^{\circ}$.
The measured forces were in good agreement with predicted 
values based on measured diffraction angles and efficiencies, with differences attributed
to multiple diffraction orders and scattering.
To our knowledge, this is the first quantitative confirmation of the equivalence of
the grating momentum and mechanical momentum of a diffraction grating.
Further, these results highlight the opportunity to replace reflective sailcraft
with diffractive sails.
To further advance the performance of diffractive sailcraft, large area single order gratings 
having a high diffraction efficiency at either a single wavelength (in the case of a
laser-driven sail) or across a broad region of the solar spectrum (in the
case of a sun-driven sail) may be developed.

\bibliographystyle{unsrt}
\bibliography{20180412_grating}

\section*{Appendix}
The equation of motion for angular displacement may be expressed
\begin{equation}
I d^2 \delta/dt^2 +{\gamma} d \delta/dt + {\kappa}\delta(t)=F R u(t-t_0)
\end{equation}
where $\gamma = 2 I \alpha$ is a damping constant, $u(t-t_0)$ is a step function, and
$t_0$ is the shutter release time.  
For small angular displacements, we assume the driving force, $F$ is a constant.
The solution of Eq. (5)  is found via Laplace transform techniques: 
$\delta (t) = \delta_1 (t) + \delta_2 (t)\ u(t-t_0)$ where

\begin{equation}
\delta_1 (t)=\mathrm{e}^{-\alpha\, t}\, \delta_0\,  \cos\!\left(\omega_1 t+\phi_0 \right)+\frac{\mathrm{e}^{-\alpha t}}{\omega_1}\left(\frac{\alpha}{2}\ \delta_0+\delta'_0 \right)
\label{delta_1}
\end{equation}

\begin{equation}
\delta_2 (t)= \frac{\mathrm{F}\, R}{\kappa}\, \left(1 - \mathrm{e}^{-\alpha\, \left(t - t_0\right)}\,  \cos\!\left(\omega_1 (t-t_0)+\phi_0 \right)\right)
\label{delta_2}
\end{equation}

\noindent where $\delta_1(t)$ is the unforced solution of Eq. (5) for $t < t_0$,
$\delta_0 = \delta (t=0)$, $\delta'_0 = d\delta /dt |_{t=0}$,  
$\omega_1=\sqrt{\omega_0^2-\alpha^2}$ is the oscillation frequency,
$\omega_0=\sqrt{\kappa/I}$ is the natural oscillation frequency, and
$\phi_0$ is the initial phase at $t=0$. 

\section*{Acknowledgement}
We thank Peter and Lihong Jansson (Hockessin, DE) for guidance and the hospitable 
use of their laboratory, and Sydor Optics (Rochester, NY) for thinning and dicing the 
diffraction grating. This research was partially supported by the National Science 
Foundation under Directorate for Engineering(ENG) (ECCS-1309517).


\end{document}